\documentclass[conference]{IEEEtran}
\IEEEoverridecommandlockouts
\usepackage{cite}
\usepackage{amsmath,amssymb,amsfonts}
\usepackage{algorithmic}
\usepackage{graphicx}
\usepackage{textcomp}
\usepackage{xcolor}
\usepackage{float}
\usepackage{mathtools}
\usepackage{latexsym}
\usepackage{amsthm}
\newcommand{\R}{\mathcal{R}}
\newcommand{\F}{\mathcal{F}}
\newcommand{\U}{\mathcal{U}}

\newcommand{\SA}{\mathcal{S}}

\def\BibTeX{{\rm B\kern-.05em{\sc i\kern-.025em b}\kern-.08em
    T\kern-.1667em\lower.7ex\hbox{E}\kern-.125emX}}
\begin{document}

\title{Approximation of Convex Envelope Using Reinforcement Learning\\
\thanks{$^{1}$The author was supported in part by a S.\ S.\ Bhatnagar Fellowship from the Council for Scientific and Industrial Research, Govt.\ of India. He thanks Prof.\ K.\ S.\ Mallikarjuna Rao for bringing to his notice the work of Oberman.}
}

\author{\IEEEauthorblockN{Vivek S. Borkar$^{1}$}
\IEEEauthorblockA{\textit{Department of Electrical Engineering} \\
\textit{Indian Institute of Technology Bombay}\\
Mumbai, India \\
borkar.vs@gmail.com}
\and
\IEEEauthorblockN{Adit Akarsh}
\IEEEauthorblockA{\textit{Department of Electrical Engineering} \\
\textit{Indian Institute of Technology Bombay}\\
Mumbai, India \\
adit.akarsh@iitb.ac.in}
}

\maketitle

\begin{abstract}
Oberman \cite{Ober1} gave a stochastic control formulation of the problem of estimating the convex envelope of a non-convex function. Based on this, we develop  a reinforcement learning scheme to approximate the convex envelope, using a variant of Q-learning for controlled optimal stopping. It shows very promising results on a standard library of test problems.
\end{abstract}


\section{Introduction}

One promising approach to non-convex optimization problems is to come up with a quick and possibly crude estimate of the convex envelope $f_c$ of the given non-convex function $f$. The convex envelope, it may be recalled, is the largest convex function whose graph lies below the graph of $f$, i.e,
$$f_c(x) = \sup_C\{g(y) : g(\cdot) \ \mbox{is convex and} \ g(y) \leq f(y) \ \forall \ y\in C\},$$
where $C\subset\R^d, d \geq 1,$ is the domain of $f$, assumed to be closed and convex. Since pointwise supremum of  a pointwise bounded family of convex functions is convex, $f_c$ is convex. Its uniqueness too is easy to verify from the above definition. Equivalently, $f_c$ is the unique function whose epigraph $\{(x,y) \in C\times\R^d : y\geq f_c(x)\}$ is the closed convex hull of the epigraph of $f$,  $\{(x,y) \in C\times\R^d : y\geq f(x)\}$ . Furthermore, it is also easy to see that $\min_Cf_c = \min_Cf$. This last fact implies that minimizing $f_c$ will yield the minimizer of $f$, so a minimization of a  good estimate of $f_c$ can be used as a surrogate for the original problem, at the least to provide a `warm start'.

The convex envelope of a function satisfies a p.d.e.\ (for `\textit{partial differential equation}'), a fact that has been analyzed extensively and made the basis of numerical schemes for estimating $f_c$ \cite{Abbasi, Carlier, Li, Ober1, Ober2, OberRuan, OberSylv, Vese}. In this work, we propose a simple off-line reinforcement learning scheme for the purpose, exploiting a connection with stochastic control, specifically controlled optimal stopping,  pointed out by Oberman in \cite{Ober1}. The scheme gives excellent results on a standard set of test functions.

The next section describes the controlled optimal stopping formulation  and the associated reinforcement learning scheme. Section 3 presents the supporting numerical experiments. Section 4 concludes with some possibilities for future work, including some theoretical issues that are left open.

\section{The learning algorithm}

\subsection{The equivalent control problem}

We shall work with $C = \R^d$ for some $d\geq 1$. Oberman observes in \cite{Ober1} that $f_c: \R^d \mapsto \R$ is the unique viscosity solution to the p.d.e.
\begin{equation}
\max\left\{u(x) - f(x), -\lambda_{min}\left(\nabla^2f(x)\right)\right\} = 0, \label{PDE}
\end{equation}
where $\lambda_{min}(\nabla^2f(x)) :=$  the least eigenvalue of  $\nabla^2f(x)$. It is observed in \cite{Ober1} that (\ref{PDE}) is the Hamilton-Jacobi-Bellman equation for the following `controlled optimal stopping' problem. Consider the $d$-dimensional controlled diffusion
\begin{equation}
dX(t) = \sqrt{2}U(t)dW(t), \ t \geq 0, \label{condiff}
\end{equation}
where $W(\cdot)$ is the standard scalar brownian motion and $U(\cdot) : [0,\infty) \mapsto S^d \ (:=$ the unit sphere in $\R^d$) is the control process. $U(\cdot)$ is assumed to be admissible in the sense that, for all $t \geq s \geq 0$, the brownian increment $W(t)-W(s)$ for $t \geq s \geq 0$  is independent of the $\sigma$-field $\F_s :=$ the completion, w.r.t.\ the underlying probability measure, of $\cap_{ s' > s}\sigma(U(y), W(y), y\leq s')$ for $t \geq 0$. (See \cite{Borkar} for  background material on controlled diffusions.) The objective is to minimize the cost functional
\begin{equation}
J(x,U(\cdot),\tau) := E[f(X(\tau))|X(0)= x] \label{cost}
\end{equation}
jointly over all admissible $U(\cdot)$ and $\{\F_t\}$-stopping times $\tau$.

\subsection{A discrete approximation}

While the traditional computational tool in this area has been to use numerical methods for p.d.e.s to solve (\ref{PDE}), we take an alternative  route. For this purpose, we discretize the control problem as follows. Fix a small $\delta > 0$ and a large integer $M >> 1$. We discretize $\R^d$ into the finite state space 
\begin{eqnarray*}
\SA &:=& \{x = [n_1\delta, n_2\delta, \cdots , n_d\delta] : \\
&& \ \ \ \ \ \ \ - \ M  \leq n_i \leq M \ \forall i, 1 \leq i \leq d\}.
\end{eqnarray*}
Also define the control space
$$\U := \{z = [z_1, \cdots , z_d] : z_i \in \{0,1\}, 1 \leq i \leq d, \ \sum_jz_j = 1\},$$
which will be our surrogate for the original control space $S^d$. Define a controlled Markov chain $X(n), n \geq 0,$  on $\SA$ as follows. For 
$x = [x_1, \cdots , x_d], y = [y_1, \cdots , y_d] \in \SA, u = [u_1, \cdots , u_d] \in \U$,
\begin{eqnarray*}
p(y|x,u) &=& \frac{1}{2} \ \mbox{if} \ y_i = x_i + \delta u_i \ \forall \ 1 \leq i \leq d, \\
        &=& \frac{1}{2} \ \mbox{if} \ y_i = x_i - \delta u_i \ \forall \ 1 \leq i \leq d, \\
&=& 0 \ \mbox{otherwise},
\end{eqnarray*}
with the forbidden transitions at the boundary, denoted $\partial_e\SA$,  replaced by a self loop at the boundary states with probability $=$ the net probability of the forbidden transitions. The controlled Markov chain property is then expressed as: for $n\geq 0$,
$$P(X(n+1) = j|X(m),U(m), m \leq n) = p(j|X(n),U(n)).$$

Let $\partial_e\SA$ denote the set of extreme points of $\SA$, i.e., its corners. 
Let $\zeta := \min\{n\geq 0: X(n)\in\partial_e\SA\}$
($=\infty$ if the set on the r.h.s.\ is empty). The cost to be minimized when this controlled Markov chain replaces the controlled diffusion is a suitably modified version of (\ref{cost}) given by 
$E[f(X(\tau\wedge\zeta))]$.
The corresponding value function is
$$V(x)  :=  \min E[f(X(\tau\wedge\zeta))| X(0)=x],$$
where the minimization is jointly over the control $\{U(n)\}$ and stopping times $\tau$ with respect to the increasing $\sigma$-fields $\F_n' := \sigma(X(m),U(m), m \leq n), n \geq 0$. Let  $\SA_0 := \SA\backslash \partial_e \SA$. The dynamic programming equation satisfied by $V(\cdot)$ is derived using standard arguments, and is given by
\begin{eqnarray}
V(x) &=& \min\Big(f(x),  \min_{u\in\U}\sum_{y \in \SA}p(y|x,u)V(y)\Big), x \in \SA_0,  \label{DP} \\
&=& \  f(x), x \in \partial_e\SA. \label{DPb}
\end{eqnarray}
$V$ need not be the unique solution to (\ref{DP})-(\ref{DPb}). 

Let $V :=$ the vector whose $i$th element is $V(i)$. Write (\ref{DP}) as $V = F(V)$ for a suitably defined $F$. Then any solution of (\ref{DP}) is a fixed point of $F$ and  an equilibrium for the iteration
\begin{equation}
V_{n+1} = F(V_n) \label{Mon}
\end{equation}
in $\R^{|\SA_0|}$ with $V_n(x) = f(x) \ \forall \ x\in\partial_e\SA$ and initial condition $V_0 \geq 0$ componentwise. This is a monotone map in the sense of \cite{Hirsch} and from Theorem 5.7 there, we know that it has a maximal and a minimal equilibrium (say, $V^*, V_*$ resp.) and all other $\omega$-limit sets of (\ref{Mon}) are sandwiched between the two. Furthermore, if $V_0 \geq V^*$ componentwise, $V_n \to V^*$. On the other hand, it is clear from the iteration that $0 \leq V_n(x) \leq f(x), x\in\SA_0$, for $n > 0$. Since the convex envelope of $f$ is the largest convex function dominated by $f$, we expect the desired solution of (\ref{DP})-(\ref{DPb}) to be the maximal equilibrium of (\ref{Mon}). This suggests that we begin with $V_0 \geq f$ componentwise.

The logic behind this specific approximation scheme is as follows. First, note that this is a two step approximation: first we truncate the state space to $[-M,M]^d$ for a suitably large $M >> 1$, and then we discretize this truncated state space. It is assumed that the global minima have reasonable sized neighbourhoods well away from the boundary of the truncated domain.
With this, while the estimate of the value function, i.e., the optimal cost, that our scheme yields, may be distorted near $\partial S$, our empirical results show that the objective of well-approximating the convex envelope is met to a very good extent away from a small neighbourhood of the boundary. 

By (\ref{DP})-(\ref{DPb}), the optimal stopping time given by
\begin{eqnarray*}
\tau^* &:=& \min\Big\{n \geq 0 : \\
&& \ \ \ \ \ \ \ \ \min_{u\in\U}\sum_{y\in\SA} p(y|X(n),u)V(y)  > f(X(n))\Big\}\wedge\zeta
\end{eqnarray*}
with $\tau^* = \infty$ in case the set on the right is empty and $\zeta=\infty$.  The optimal $U(n), 0\leq n < \tau^*,$ is given by
$U(n) = argmin\left(\sum_{y\in\SA}p(y|X(n), \cdot)\right)$ (with ties resolved according to any fixed tie-breaking protocol). The details are omitted as they are standard. The possibility that $\tau^*, \zeta=\infty$ cannot, however, be ruled out. We comment on this later. 

This framework facilitates using a Q-learning algorithm for the problem, which we discuss next.

\subsection{Q-learning scheme}

Define an additional control variable $z \in \{0,1\}$ corresponding to `\textit{stop}' (for $z = 0$) and `\textit{continue}' (for $z=1$). Then (\ref{DP}) can be rewritten as
$$V(x) = \min_{u,z}\left(z\sum_{y\in\SA}p(y|x,u)V(y) + (1-z)f(x)\right)$$
for $x\in\SA_0$, with $V(x) = f(x)$, $x \in \partial_e\SA$. Define the corresponding Q-value as
$$Q(x,u,z) = z\left(\sum_{y\in\SA}p(y|x,u)V(y)\right) + (1-z)f(x), \  x \in \SA_0,$$
with $Q(x,u,z) = f(x), x \in \partial_e\SA$. It is seen to satisfy the equation
\begin{eqnarray}
Q(x,u,z) &=& z\left(\sum_{y\in\SA}p(y|x,u)\min_{u',z'}Q(y,u',z')\right) \nonumber \\
&&+ \ (1-z)f(x), \ x\in\SA_0, \label{Q-DP}
\end{eqnarray}
with $Q(x,u,z) = f(x)$ for $x\in\partial_e\SA$. The associated `Q-value iteration' to solve this equation is
\begin{eqnarray}
Q_{n+1}(x,u,z) &=& z\left(\sum_{y\in\SA}p(y|x,u)\min_{u',z'}Q_n(y,u',z')\right) \nonumber \\
&& + \ (1-z)f(x), \ x \in \SA_0, \label{Q-VI}
\end{eqnarray}
with $Q_n(x,u,z) = f(x), x \in \partial_e\SA$. Routine arguments show that  the equation (\ref{Q-DP}) is of the form $Q = F'(Q)$ for a map $F' : \R^{2|\SA_0|\times|\U|} \mapsto \R^{2|\SA_0|\times|\U|}$ which is non-expansive w.r.t.\ the  $\|\cdot\|_\infty$ norm, with a fixed point given by the desired Q-value, i.e., the solution to (\ref{Q-DP}).

Following the standard derivation of the Q-learning algorithm for discounted cost \cite{Watkins}, we write the Q-learning algorithm for our problem as the stochastic approximation scheme for solving (\ref{Q-DP}). That is, we replace the conditional expectation in (\ref{Q-VI}) by evaluation at a simulated transition according to the same conditional probability, and then make an incremental move to a convex combination of the previous iterate and the suggested correction, with a slowly decreasing weight on the latter. This leads to 
\begin{eqnarray}
\lefteqn{Q_{n+1}(x,u,z) = (1 - a(n))Q_n(x,u,z) +} \nonumber \\
&&a(n)\left(z\min_{u',z'}Q_n(X'(n+1),u',z') + (1-z)f(x)\right) \nonumber \\
&=& Q_n(x,u,z) + a(n)\Big(z\min_{u',z'}Q_n(X'(n+1),u',z') + \nonumber \\
&& (1-z)f(x) - Q_n(x,u,z)\Big), \label{Qlearn}
\end{eqnarray}
where $X'(n+1)$ is simulated according to the law $p(\cdot|x,u)$. The asynchronous counterpart of this scheme based on a single run of a controlled Markov chain $\{(X(n),U(n),Z(n))\}$ where $\{Z(n)\}$ is the $\{0,1\}$-valued control sequence, is:
\begin{eqnarray}
\lefteqn{Q_{n+1}(x,u,z) = Q_n(x,u,z) \ + } \nonumber \\
&&\ \ \ \ \ \ \ \ a(n)I\{X(n) = x, U(n) = u, Z(n) = z\}\times \nonumber \\
&&\Big(z\min_{u',z'}Q_n(X(n+1),u',z') + \nonumber \\
&& \ \ \ \ \ \ \ \ (1-z)f(x) - Q_n(x,u,z)\Big). \label{Qlearn2}
\end{eqnarray}
Here $I\{\cdots\}$ denotes the indicator function, $= 1$ if `$\cdots$' holds, $=0$ if not. By standard results from the `o.d.e.\ approach to stochastic approximation' (see, e.g., \cite{BorkarBook}, Chapter 2), the iterates track the the asymptotic behaviour of differential equation 
$$\dot{Q}(t) = F'(Q(t)) - Q(t).$$
From the theory of cooperative differential equations, if applicable, one would expect the solution $Q(t)$ of this equation to converge to the maximal fixed point if initiated in a state that dominates it. The catch here is that the available theory for monotone dynamics in continuous time is only for differential equations driven by smooth vector fields, mere Lipschitz continuity is not enough. Nevertheless, it gives us reasons to believe that the iterates will converge to the desired maximal equilibrium if initiated as prescribed above, with very high probability. In our numerical experiments, it was always so.
Starting with low initial Q-values (e.g., zero) can lead to wrong equilibrium that is not maximal, a fact confirmed by our numerical experiments. Hence we always start with an initial condition that exceeds $\|f\|_\infty$.  We also found that fixing $V(x) = f(x)$ in this manner, but only at the corners of the rectangular domains, suffices, so we did so.

\section{Numerical experiments}
We applied the  scheme to some one or two dimensional functions from a standard library of test functions for non-convex optimization, with the following hyperparameter choices:
\begin{eqnarray}
\nonumber&&Q_0(x,u,z) = L\ \forall\ x \in\SA_0, u \in \U, z \in \{0,1\}\\ 
\nonumber&&Q_0(x,u,z) = f(x) \forall\ x \in \partial_e\SA, u \in \U, z \in \{0,1\}\\ 
\nonumber &&a(n) = \frac{1}{[1+\frac{n}{1000000}]}\ n \in \mathbb{N}
\end{eqnarray}
Here, $L$ is any constant greater than the function value at all points in the domain.

\begin{figure}[H]
    \centering
    \includegraphics[width = 0.4\textwidth]{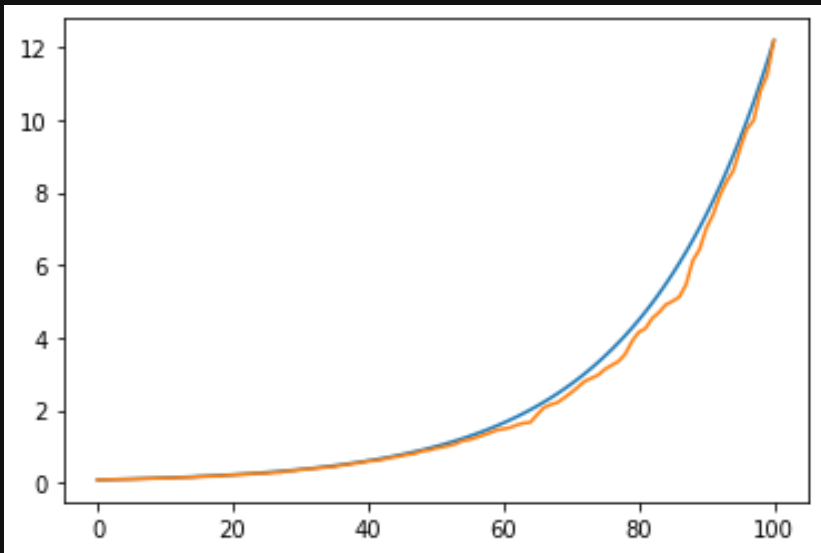}
    \caption{$f(x) = \exp{x}$; orange is the obtained envelope, blue is the original function}
    \label{fig:my_label}
\end{figure}

For the dropwave function\cite{website},
$$f(x,y) = -\frac{1 + cos(12\sqrt{x^2+y^2)}}{0.5(x^2+y^2)+2}$$

\begin{figure}[H]
    \centering
    \includegraphics[width = 0.4\textwidth]{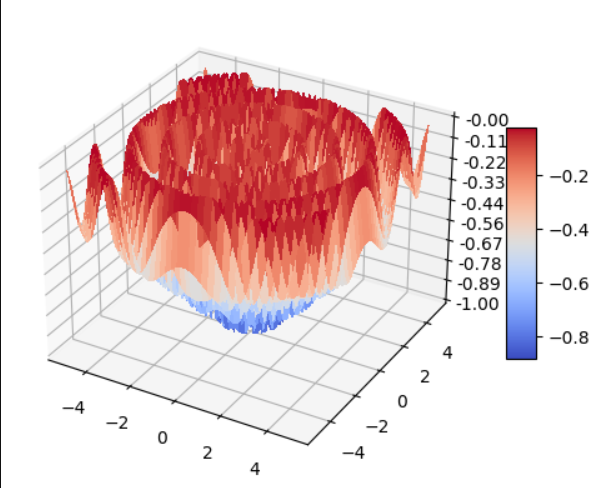}
    \caption{Dropwave function}
    \label{fig:my_label}
\end{figure}

\begin{figure}[H]
    \centering
    \includegraphics[width = 0.4\textwidth]{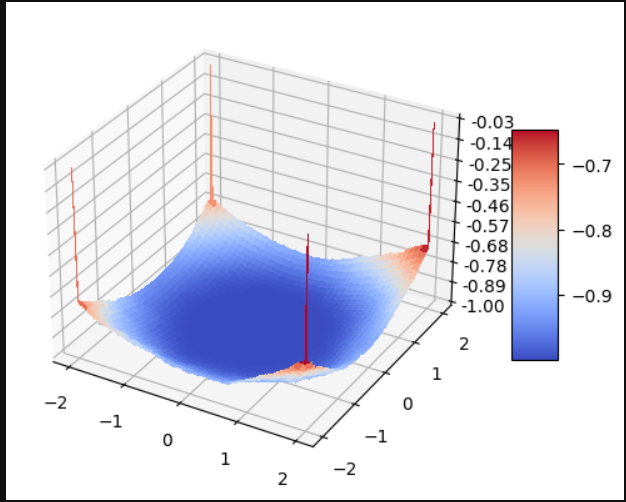}
    \caption{Obtained convex envelope for dropwave function}
    \label{fig:my_label}
\end{figure}

For the $sinc$ function, $f(x,y) = \frac{sin(\sqrt{x^2+y^2})}{\sqrt{x^2+y^2}}$,
\begin{figure}[H]
    \centering
    \includegraphics[width = 0.4\textwidth]{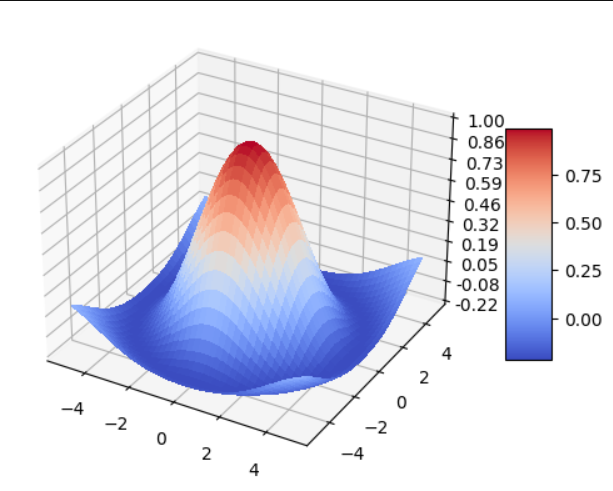}
  \caption{Sinc function}
    \label{fig:my_label}
\end{figure}

\begin{figure}[H]
    \centering
    \includegraphics[width = 0.4\textwidth]{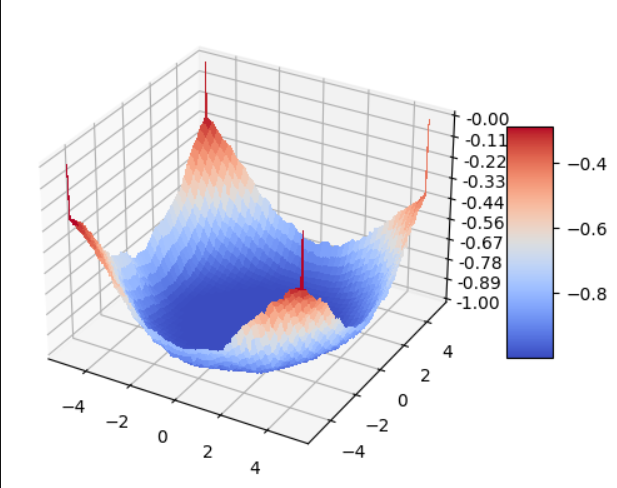}
    \caption{Obtained convex envelope for sinc function}
    \label{fig:my_label}
\end{figure}

For the Ackley function,
$f(x,y) = -a \exp{\left(-b\sqrt{\frac{1}{2} (x^2+y^2)}\right)} - exp{\left(\frac{1}{2} (cos(cx) + cos(cy))\right)} +a + \exp{(1)}$

\begin{figure}[H]
    \centering
    \includegraphics[width = 0.4\textwidth]{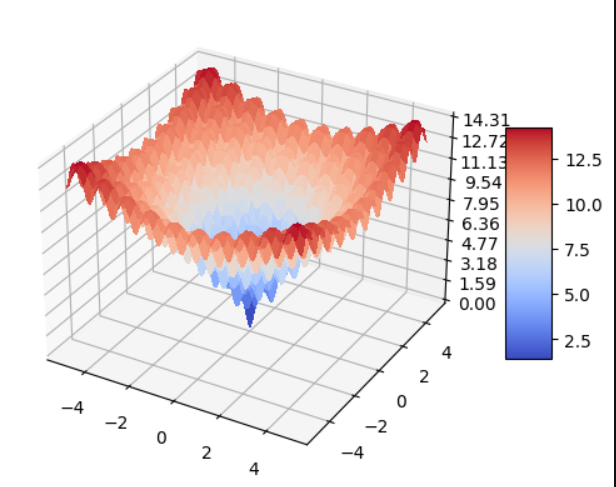}
  \caption{Ackley function}
    \label{fig:my_label}
\end{figure}

\begin{figure}[H]
    \centering
    \includegraphics[width = 0.4\textwidth]{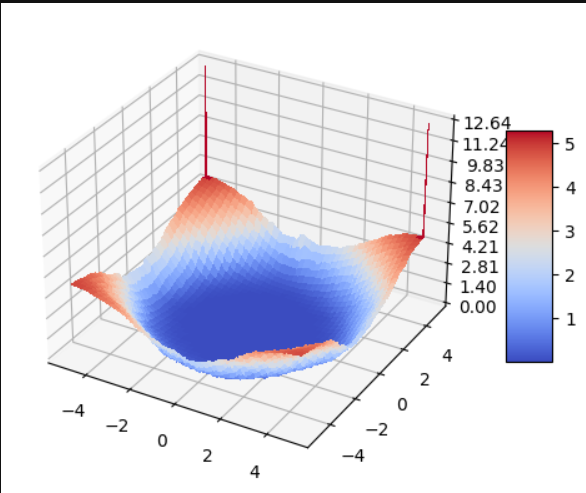}
  \caption{Obtained convex envelope for Ackley function}
    \label{fig:my_label}
\end{figure}

For the Levy function, given by
$f(x,y)=\sin ^2\left(\pi w_1\right)+\left(w_1-1\right)^2\left[1+10 \sin ^2\left(\pi w_1+1\right)\right]+\left(w_2-1\right)^2\left[1+\sin ^2\left(2 \pi w_2\right)\right]$, where $w_1=1+\frac{x-1}{4}$ and $w_2=1+\frac{y-1}{4}$

\begin{figure}[H]
    \centering
    \includegraphics[width = 0.4\textwidth]{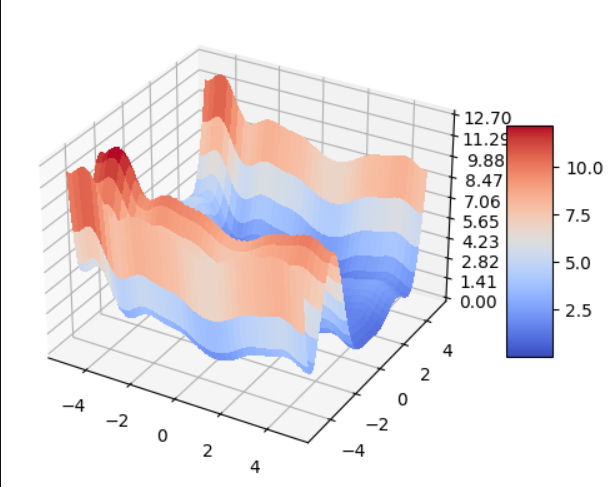}
  \caption{Levy function}
    \label{fig:my_label}
\end{figure}
\begin{figure}[H]
    \centering
    \includegraphics[width = 0.4\textwidth]{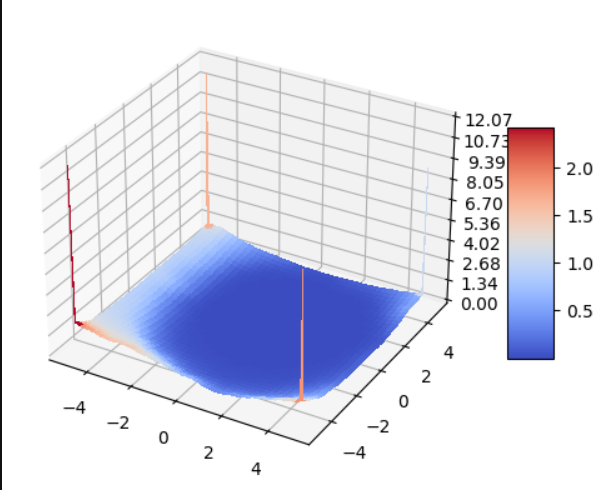}
  \caption{Obtained convex envelope for Levy function}
    \label{fig:my_label}
\end{figure}

For the origin-shifted Easom function, given by
$f(x,y)=-cos(x+\pi)cos(y+\pi)exp(-x^2-y^2)$

\begin{figure}[H]
    \centering
    \includegraphics[width = 0.4\textwidth]{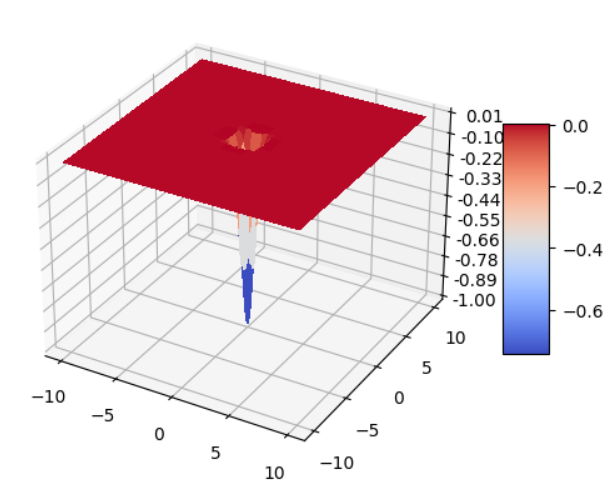}
  \caption{Easom function}
    \label{fig:my_label}
\end{figure}

\begin{figure}[H]
    \centering
    \includegraphics[width = 0.4\textwidth]{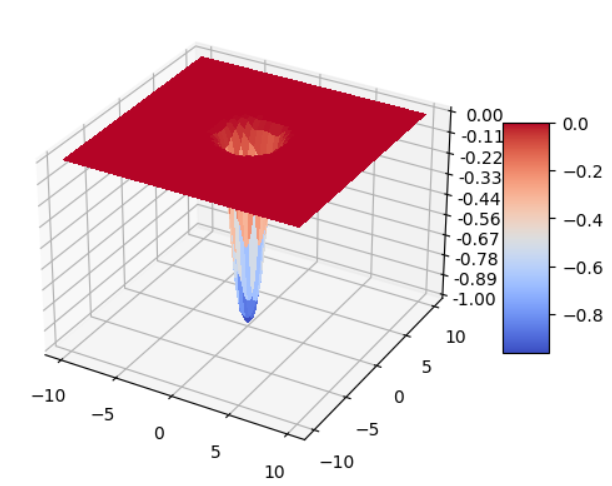}
  \caption{Obtained convex envelope for Easom function}
    \label{fig:my_label}
\end{figure}

For the Rastrigin function, given by
$f(x,y)=20+ x^2+ y^2 -10cos(2\pi x) - 10 cos (2 \pi y)$

\begin{figure}[H]
    \centering
    \includegraphics[width = 0.4\textwidth]{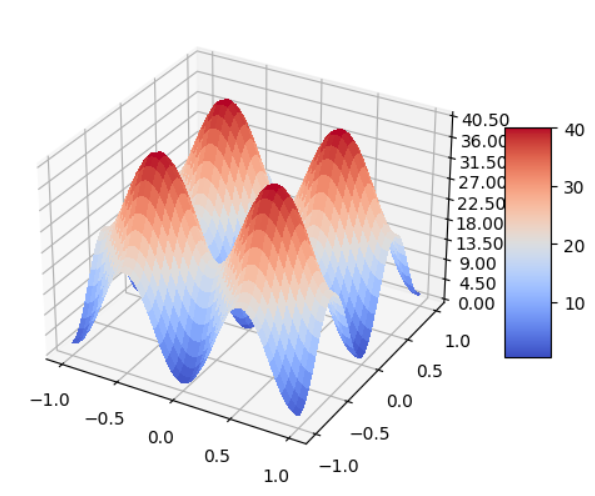}
  \caption{Rastrigin function}
    \label{fig:my_label}
\end{figure}

\begin{figure}[H]
    \centering
    \includegraphics[width = 0.4\textwidth]{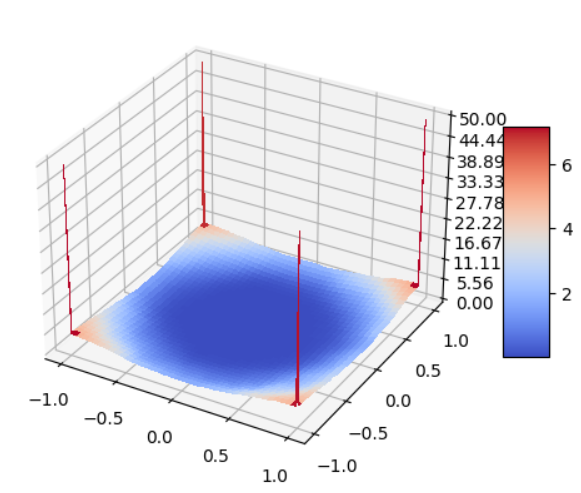}
  \caption{Obtained convex envelope for Rastrigin function}
    \label{fig:my_label}
\end{figure}

For the Schubert function, given by
$f(x,y) = \left(\sum_{i=1}^5i cos((i+1)x+i\right)\left(\sum_{i=1}^5i cos((i+1)y+i\right)$

\begin{figure}[H]
    \centering
    \includegraphics[width = 0.4\textwidth]{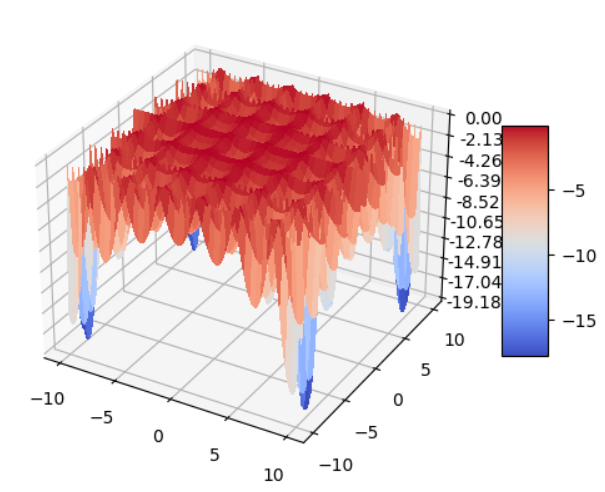}
  \caption{Schubert function}
    \label{fig:my_label}
\end{figure}

\begin{figure}[H]
    \centering
    \includegraphics[width = 0.4\textwidth]{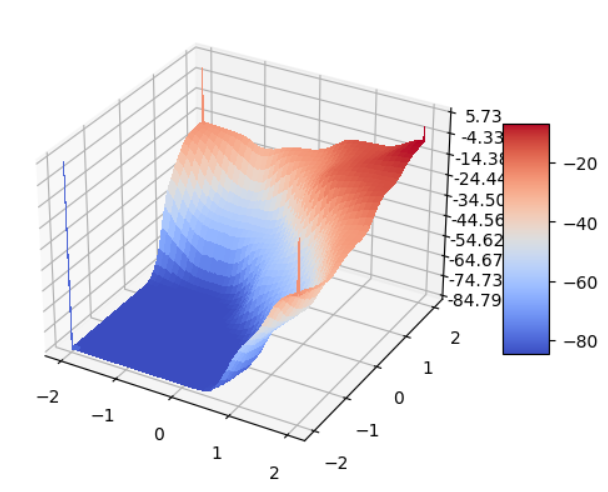}
  \caption{Obtained convex envelope for Schubert function}
    \label{fig:my_label}
\end{figure}

For the Holder Table function, given by
$f(x,y) =-\left|\sin \left(x\right) \cos \left(y\right) \exp \left(\left|1-\frac{\sqrt{x^2+y^2}}{\pi}\right|\right)\right|$

\begin{figure}[H]
    \centering
    \includegraphics[width = 0.4\textwidth]{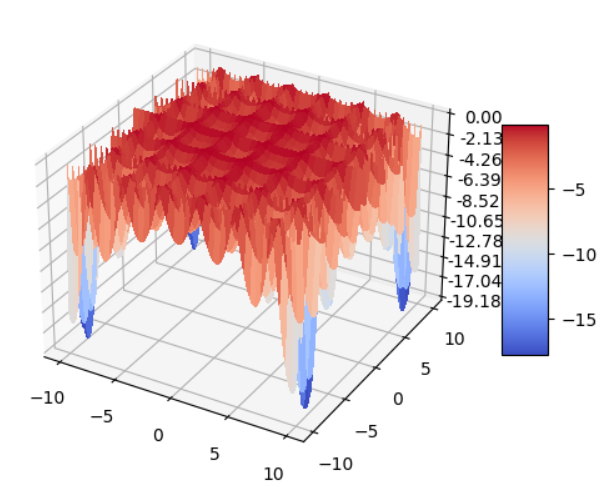}
  \caption{Holder Table function}
    \label{fig:my_label}
\end{figure}

\begin{figure}[H]
    \centering
    \includegraphics[width = 0.4\textwidth]{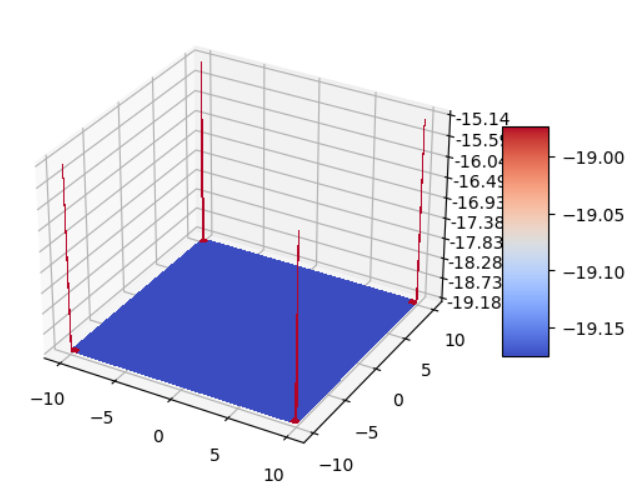}
  \caption{Obtained convex envelope for Holder Table function}
    \label{fig:my_label}
\end{figure}

\section{Future directions}

While success on the much used family of test functions for convex optimization does serve as a proof of concept, one needs to see how the scheme will scale with dimensionality. In fact for high dimensions, one may have to resort to function approximation techniques for the Q-values. This is an important possibility that needs to be checked. Given the eventual objective of optimizing $f$, it is also of interest to see whether such approximations can be embedded in an overall scheme that also optimizes $f_c$. On the theoretical side, we need a counterpart of the existing theory of monotone dynamical systems in continuous time, specifically, the so called cooperative differential equations, to those driven by Lipschitz vector fields. In addition,  one needs to prove a rigorous functional central limit theorem justifying the controlled Markov chain approximation of the controlled diffusion (see, e.g., \cite{Kushner}). Also, error estimates for this approximation would be of great interest.


\begin{thebibliography}{99}

\bibitem{Abbasi} Abbasi, B.\ and Oberman, A., 2018. Computing the level set convex hull. Journal of Scientific Computing, 75, 26-42.


\bibitem{Borkar} Borkar, V., 2005. Controlled diffusion processes. Probability Surveys, 213-244.

\bibitem{BorkarBook} Borkar, V., 2022. Stochastic Approximation: A Dynamical Systems Viewpoint, Hindustan Publishing Agency and Springer Nature.

\bibitem{Carlier} Carlier, G.\ and Galichon, A., 2012. Exponential convergence for a convexifying equation. ESAIM: Control, Optimisation and Calculus of Variations, 18(3), 611-620.

\bibitem{Kushner} Kushner, H.\ and Dupuis, P., 2000. Numerical Methods for Stochastic Control Problems in Continuous Time (2nd edition), Springer.

\bibitem{Hirsch} Hirsch, M.\ W.\ and Smith, H., 2005. Monotone maps: a review. Journal of Difference Equations and Applications, 11(4-5), 379-398.

\bibitem{Li} Li, W.\ and Nochetto, R., 2022. Two-scale methods for convex envelopes. Mathematics of Computation, 91(333), 111-139.

\bibitem{Ober1} Oberman, A., 2007. The convex envelope is the solution of a nonlinear obstacle problem. Proceedings of the American Mathematical Society, 135(6), 1689-1694.

\bibitem{Ober2} Oberman, A., 2008. Computing the convex envelope using a nonlinear partial differential equation. Mathematical Models and Methods in Applied Sciences, 18(05), 759-780.

\bibitem{OberRuan} Oberman, A.\  and Ruan, Y., 2017. A partial differential equation for the rank one convex envelope. Archive for Rational Mechanics and Analysis, 224, 955-984.

\bibitem{OberSylv} Oberman, A.\ and Silvestre, L., 2011. The Dirichlet problem for the convex envelope. Transactions of the American Mathematical Society, 363(11), 5871-5886.


\bibitem{Vese} Vese, L., 1999. A method to convexify functions via curve evolution. Communications in Partial Differential Equations, 24(9-10), 1573-1591.

\bibitem{Watkins} Watkins, C., 1989. Learning from delayed rewards. Ph.D.\ Thesis, King's College, University of Cambridge.

\bibitem{website} S. Surjanovic and D. Bingham, “Virtual library of simulation experiments:” Optimization Test Functions and Datasets. [Online]. Available at: https://www.sfu.ca/~ssurjano/optimization.html. [Accessed: 28-Apr-2023]. 

\end{thebibliography}
\end{document}